\documentclass[aps,prl,reprint,showpacs,superscriptaddress]{revtex4-1}
\usepackage[english]{babel}
\usepackage{amsmath,amssymb,bbm,graphicx,color,comment,txfonts}
\usepackage[bookmarks=true,colorlinks,citecolor=blue,urlcolor=blue]{hyperref}
\usepackage{dsfont}
\usepackage[normalem]{ulem}
\usepackage{physics}
\usepackage{physics}
\usepackage{tikz}
\usepackage{mathdots}
\usepackage{yhmath}
\usepackage{cancel}
\usepackage{color}
\usepackage{siunitx}
\usepackage{array}
\usepackage{multirow}
\usepackage{amssymb}
\usepackage{gensymb}
\usepackage{tabularx}
\usepackage{extarrows}
\usepackage{booktabs}
\usetikzlibrary{fadings}
\usetikzlibrary{patterns}
\usetikzlibrary{shadows.blur}
\usetikzlibrary{shapes}

\usepackage{orcidlink}

\begin{document}

\date{\today}

\title{Quantum Mpemba effect in chaotic systems with conservation laws}

\author{Thomas Martin M\"uller~\orcidlink{0000-0002-6924-6741}}
\affiliation{The Abdus Salam International Centre for Theoretical Physics (ICTP), Strada Costiera 11, 34151 Trieste, Italy}
\author{Silvia Pappalardi~\orcidlink{0000-0001-6931-8736}}
\affiliation{Institut f\"ur Theoretische Physik, Universit\"at zu K\"oln, Z\"ulpicher Straße 77, 50937 K\"oln, Germany}
\author{Rosario Fazio~\orcidlink{0000-0002-7793-179X}}
\affiliation{The Abdus Salam International Centre for Theoretical Physics (ICTP), Strada Costiera 11, 34151 Trieste, Italy}
\affiliation{Dipartimento di Fisica “E. Pancini”, Università di Napoli “Federico II”, Monte S. Angelo, I-80126 Napoli, Italy}

\begin{abstract}
Closed chaotic quantum systems relax after a quench into a Gibbs ensemble. At late times, the relaxation speed is determined by their conservation laws and hydrodynamics. 
As a result, there exist pairs of initial states which thermalize to the same ensemble, yet exhibit drastically different hydrodynamic relaxation. 
We show in two chaotic spin chains how this enables a simple and robust realization of the quantum Mpemba effect: a system initially closer to equilibrium relaxes slower than one that starts farther away, despite both approaching the same final state.
\end{abstract}

\maketitle
\begin{figure}
    \centering
    \includegraphics[width=1\linewidth]{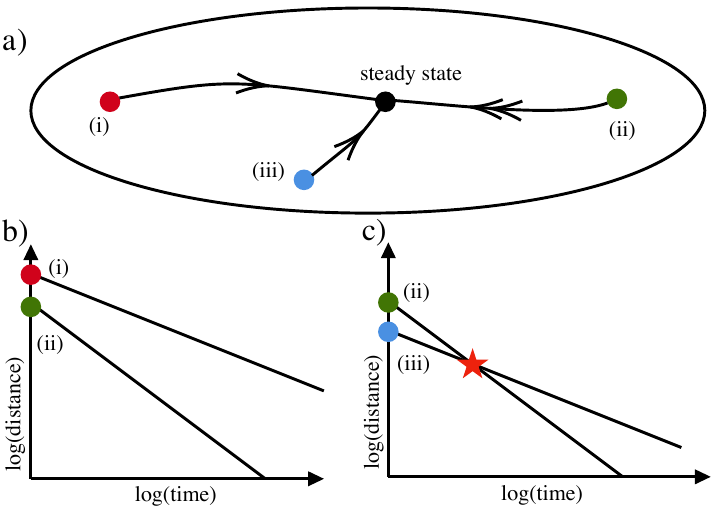}
    \caption{Mpemba effect in the hydrodynamic regime. a) Sketch of the
relaxation from different initial states (i)-(iii) into a unique stationary state in Hilbert space. Even though all initial states relax into the same state, they may have drastically different relaxation speeds, marked by the arrows. In b), c) we sketch the corresponding relaxation of a witness of the distance from the stationary state to zero in the hydrodynamic regime. b) If an initial state (i) further from equilibrium relaxes slower than one closer (ii), there is no crossing and no Mpemba effect. c) If another state (iii) closer than (ii) to equilibrium also relaxes more slowly, this gives rise to a crossing marked by the star.}
    \label{fig:1}
\end{figure}

Originally, the Mpemba effect refers to the counterintuitive observation that an initially hot sample of water freezes quicker than an initially colder one~\cite{Mpemba_1969}, i.e. closer to the freezing point. More recently, a closely related phenomenon has been discussed, where a classical~\cite{Lu_2017}, or a quantum system~\cite{Carollo_2021,Ares_2023} initialized further -- according to an appropriately chosen measure~\cite{Summer_2025} -- from its stationary state can relax faster than one initialized in closer proximity to the final steady state, see Fig.~\ref{fig:1}a). If this occurs in a quantum system, we call it \textit{quantum Mpemba effect} (QME)~\cite{Ares_2025,Yu_2025_review}. Systems where the QME was found include closed integrable~\cite{Ares_2023,Murciano_2024,Yamashika_2024,Rylands_2024,Ares_2025,Ares_2025_mixed,Ares_2025_simpler,Caceffo_2024,Chalas_2024,Yamashika_2025,Calabrese_2026}, chaotic~\cite{Turkeshi_2024,Qian_2025,Li_2025,Aditya_2025,Bhore_2025,Yu_2025,Alishahiha_2025,Hallam_2025,Yamashika_2026,Ulcakar_2025,Yamashika_2026_resonances,Liu_2024_random,Li_2025_Circuit} and many-body localizing~\cite{Liu_2025} quantum systems as well as open~\cite{Carollo_2021,Nava_2024,Nava_2025,Zatsarynna_2025,Chatterjee_2023,Chatterjee_2024,Chatterjee_2025,Moroder_2024,Qian_2025,Medina_2025,Dong_2025,Li_2025,Furtado_2025,Longhi_2025,Liu_2024,Zhao_2025,Strachan_2025,Wei_2025,Chattopadhyay_2026,Das_2025,Bao_2026}, and monitored dynamics~\cite{Di_Giulio_2025}. The QME has been experimentally observed in quenches in synthetic quantum matter~\cite{Joshi_2024,Zhang_2025,Xu_2025,Chatterjee_2025}. 

The underlying mechanism determining the QME is tightly bound to the specific dynamics that governs the quantum system. In a Lindbladian setting with few degrees of freedom, the QME can be constructed by tuning the overlap of the initial state with the slowest decaying mode to zero. Then, the leading behavior at late times is governed by the second-slowest decaying mode, letting the system relax exponentially faster than a generic initial state~\cite{Carollo_2021}. In integrable models, typically one can analytically solve the system using a mapping to quasi-particles and predict the QME in a measure quantifying the restoration of a symmetry, the entanglement asymmetry~\cite{Ares_2023}. For chaotic many-body systems, the QME has been seen numerically~\cite{Ares_2023,Yu_2025,Alishahiha_2025,Hallam_2025,Turkeshi_2024,Qian_2025,Li_2025}, but a full physical explanation is still lacking. Filling this gap is the goal of this work. 

\begin{figure}
    \centering
    \includegraphics[width=\linewidth]{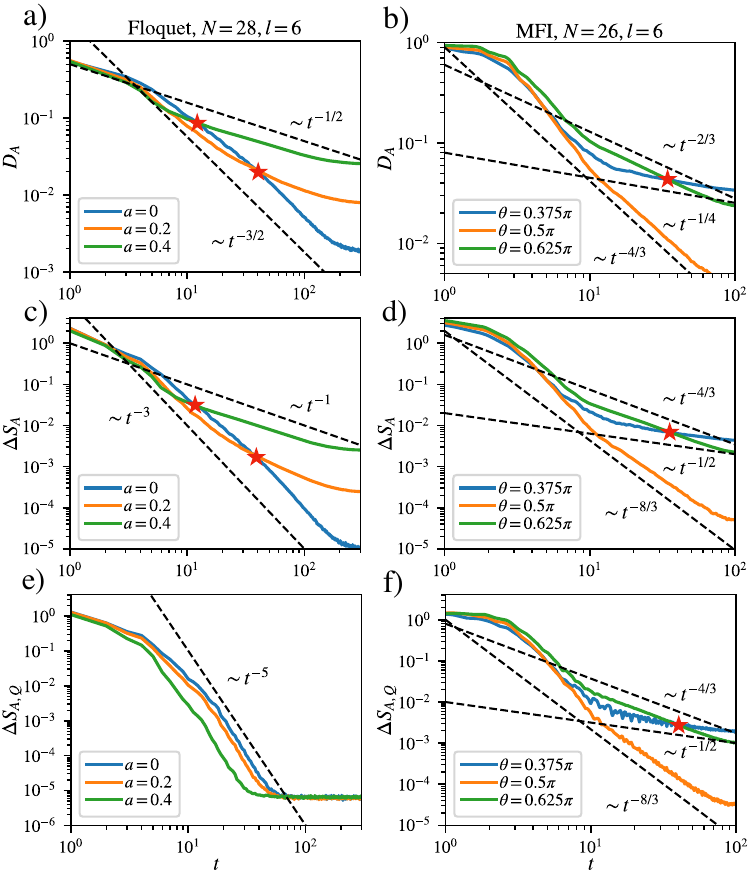}
    \caption{a),b) QME due to different power-laws occurring in the hydrodynamic regime, witnessed by the trace distance from the infinite temperature state for the Floquet (a) and the MFI (b) model. c), d) The entanglement entropy approaches the maximum $l \log 2$ according to various different power laws for different initial states in both the Floquet (c) and the MFI model (d). We find Mpemba crossings in both models, marked by stars. e) In the Floquet model, the magnetization is conserved and therefore the slowest decaying hydrodynamic modes are all symmetric. Therefore they do not appear in the entanglement asymmetry and it shows accelerated relaxation. By hydrodynamics arguments, slower relaxation $\sim t^{-3}$ is expected to be seen for $a \neq 0$ at late times, but we always find $~\sim t^{-5}$. We suspect this to be due to strong finite size effects for this quantity. f) In the MFI model, $\hat{Q}$ is not conserved such that the asymmetry is not related to the slowest decaying modes. Therefore we essentially find the same dynamics as for the entanglement entropy, including an Mpemba crossing. Dashed lines are guides to the eye indicating the various power-law behaviors at intermediate times.}
    \label{fig:2}
\end{figure}

In this Letter, we present a general mechanism on how some specific initial states undergo faster thermalization than a generic initial state, which explains the QME in chaotic many-body systems. We study systems with a conservation law, where thermalization occurs algebraically instead of exponentially, due to hydrodynamic long-time tails~\cite{Mukerjee_2006, Lux_2014,delacretaz2020heavy, Matthies_2026}. The appearing power-laws can be modified for specific initial states, where the fluctuations of initial and stationary state are in agreement \cite{Matthies_2026}, which results in the robust and systematic appearance of the QME in the hydrodynamic regime, as shown in Fig.~\ref{fig:2}. Our numerical observations will be further corroborated by analytic results using a classical hydrodynamic description~\cite{Matthies_2026}.

\textit{General setting} --  The Mpemba effects rests on three critical ingredients: (I) A system setup and generator of the dynamics. In our case, this is a spin chain undergoing chaotic symmetric Floquet or Hamiltonian dynamics. (II) Different initial states, which relax under the given dynamics into a stationary state. An optional additional constraint that we find important is to require that the various initial states locally relax into the same unique stationary state. This is typically not taken into consideration in earlier work on unitarily evolving systems~\cite{Murciano_2024,Yamashika_2024,Rylands_2024,Ares_2025,Ares_2025_mixed,Ares_2025_simpler,Caceffo_2024,Chalas_2024,Yamashika_2025,Ares_2023,Yu_2025,Alishahiha_2025,Hallam_2025,Turkeshi_2024,Qian_2025,Li_2025}. In this work, we always pick initial states that locally relax into the infinite temperature state. (III) A (local or quasi-local) measure or witness that quantifies the distance between the instantaneous state of the system and the stationary state. Commonly used distance quantifiers are the trace distance, fidelity~\cite{Nielsen_Chuang_2010} or entanglement asymmetry~\cite{Ares_2023}, which we check in our numerical simulations. 

If the distance witness of one state is initially larger than the one of the other state, but there is a time $t_\text{M}$ at which this order swaps, we call this QME (see Fig.~\ref{fig:1}). It turns out that this scenario is realized whenever the speed at which relaxation occurs in the chosen distance measure depends on the initial state. Hence, in order to observe the Mpemba effect, we need to find situations where two initial states relax into the same stationary state under the same dynamics at qualitatively different speed.

\textit{Models} -- We consider two different cases of chaotic dynamics of a spin-$\frac{1}{2}$ chain described by a (pure) state $\ket{\psi}$. In both models, we consider periodic boundary conditions in a system of $N$ sites, $i=1 \dots N$, and are interested in the thermodynamic limit $N \to \infty$. \\
\noindent
1) $\mathrm{U}(1)$ symmetric Floquet dynamics~\cite{Matthies_2026} -- The state after $n$ periods~\footnote{We identify the discretized time as $t=t_n=n$.} evolves according to $\ket{\psi(t_{n+1})}=\hat{U}_\text{FL} \ket{\psi(t_n)}$ with
\begin{equation}
    \hat U_\text{FL}=e^{-i\hat{H}_\gamma} e^{-i\hat{H}_\beta^\text{e}} e^{-i\hat{H}_\beta^\text{o}} e^{-i\hat{H}_\alpha},
\end{equation}
and $\hat{H}_\alpha=-\alpha\sum_{i} \hat{Z}_i \hat{Z}_{i+1}$, $\hat{H}_\beta^\text{e/o}=-\beta \sum_{i \text{ even/odd}} (\hat{X}_i \hat{X}_{i+1} + \hat{Y}_i \hat{Y}_{i+1})$, $\hat{H}_\gamma=-\gamma\sum_{i} \hat{Z}_i \hat{Z}_{i+2}$~\footnote{We choose the parameters $\alpha=2,\beta=0.25,\gamma=1$ \cite{Matthies_2026}}.  \\
\noindent
2) Mixed field Ising (MFI) model~\cite{Bhore_2025,Kim_2013,Atas_2015,Banuls_2011, maceira2025thermalization} -- We evolve the state according to the Hamiltonian
\begin{equation}
    \hat{H}_\text{MFI}=\sum_i (J\hat{Z}_i \hat{Z}_{i+1}+h_x \hat{X}_i+h_z \hat{Z}_i).
\end{equation}
We take $J=1$, $h_x=(5+\sqrt{5})/8$, $h_z=(1+\sqrt{5})/4$. 

\begin{figure}
    \centering
    \includegraphics[width=\linewidth]{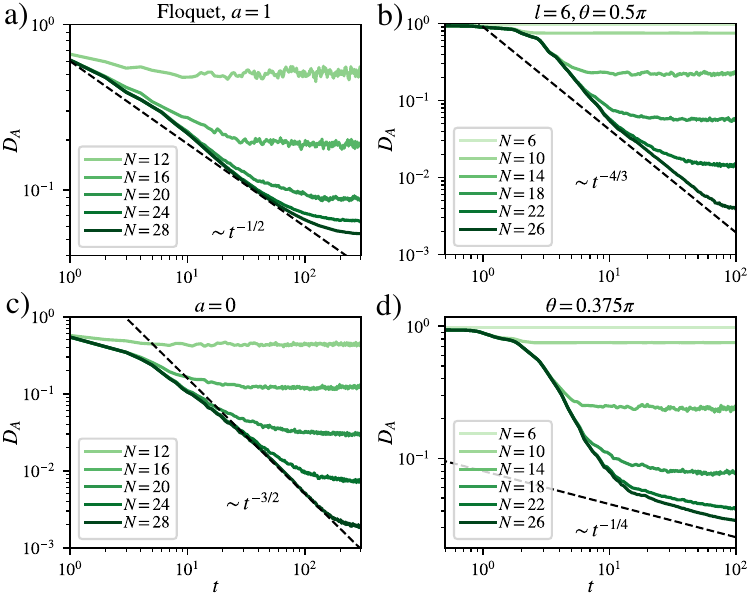}
    \caption{Emergence of power-law relaxation in the hydrodynamic regime and the effect of finite system size. For all considered initial states, we find thermalization into $\hat{\Lambda}_A \propto \hat{1}$ upon increasing system size $N$ and time $t$. In the Floquet system (a),c)), we observe power-laws at intermediate times that are compatible with hydrodynamic predictions, with substantial sub-leading corrections~\cite{Matthies_2026}. In particular, for $a=0$, we find accelerated relaxation due to an additional symmetry of the initial state. b),d) In the MFI model, clearly algebraic relaxation emerges, with unconventional exponents depending on the specific initial state. We show data for $l=6$ in all plots.}
    \label{fig:3}
\end{figure}

Both models have a single conserved quantity $\hat{Q}=\sum_i \hat{Q}_{i}$, where $[\hat{Q},\hat{U}]=0$. In the Floquet model, $\hat{Q}=\sum_i\hat{Z}_i$, the magnetization is conserved. The MFI model conserves energy, $\hat{Q}=\hat{H}_\text{MFI}$. Then, by the eigenstate thermalization hypothesis~\cite{Deutsch_2018}, they are both expected to evolve into a state that is locally indistinguishable from a Gibbs state fixed by the expectation value of the respective conserved quantity $\hat{Q}$ in the initial state. Hence, if we pick a connected subsystem $A$ of finite length $l$, we expect that
\begin{equation}
    \lim_{t \to \infty} \lim_{N\to \infty} \hat{\rho}_{A}(t)
= \lim_{N \to \infty} \hat\Lambda_A,
\end{equation}
where $\hat{\rho}_{A}(t)=\Tr_B \ket{\psi(t)} \bra{\psi(t)}$ is the reduced density matrix of the system at time $t$, $\hat{\Lambda}_A=\Tr_B \hat{\Lambda}$ is the reduced density matrix of the Gibbs state $\hat{\Lambda}=e^{-\lambda \hat{Q}}/\Tr e^{-\lambda \hat{Q}}$ and $B$ is the rest of the system. $\lambda$ needs to be fixed to reproduce the conserved quantity correctly, i.e. $\langle \hat{Q} \rangle_0=\Tr \hat{Q} \hat{\Lambda}$~\footnote{We introduce the notation $\bra{\psi(t)} \dots \ket{\psi(t)}=\langle \dots \rangle_t$}. To compare the relaxation dynamics into a unique stationary state from different initial states, we focus on initial states where $\langle \hat{Q} \rangle_0=0$ such that $\lambda = 0$ and therefore $\hat{\Lambda} \sim \hat{1}$ is the infinite temperature state. In finite systems, $\hat{\rho}_A(t\to \infty) - \hat{\Lambda}_A$ 
has an offset that vanishes as $N\to \infty$, c.f. Fig.~\ref{fig:2},~\ref{fig:3}.

There is an exponentially large number of initial states that thermalize at infinite temperature. For the sake of clarity we consider a set of specific initial states and analyze their speed of thermalization. Therefore, we limit the discussion to initial product states $\ket{\psi(0)}=\bigotimes_i \ket{\psi_0^i}$ with a simple translational symmetry. On each site, we parametrize the state as 
\begin{equation}
    \ket{\psi_0^i} = \cos \frac{\theta_i}{2} \ket{0} + e^{i\phi_i} \sin \frac{\theta_i}{2} \ket{1},
\end{equation}
For the Floquet model, we take $\phi_i=0$ and $\cos \frac{\theta_i}{2}= \frac{1+(-1)^i a}{2} $ where $a\in [0,1]$. For $a=0$, the initial state is $+$-polarized and translation-invariant, i.e. $\langle \hat{Z}_i \rangle_0=0$ everywhere. For $a>0$, we break this initial symmetry and have $\langle \hat{Z}_i \rangle_0=(-1)^i a$, a staggered magnetization which vanishes upon a spatial average. $a$ therefore parametrizes a series of different initial states that obey $\langle \hat{Q} \rangle_0=\sum_i \langle \hat{Z}_i \rangle_0=0$ and therefore relax into the infinite temperature state. In the MFI model, we choose $\phi_i=\phi,\theta_i=\theta$ equal on all sites, with the constraint that $\langle \hat{Q} \rangle_0=\langle \hat{H}_\text{MFI} \rangle_0=0$, satisfied for $\phi=\arccos((-h_z- \cos \theta)/h_x \tan \theta)$. This ensures that all states parametrized by $\theta$ thermalize into the infinite temperature state under the MFI dynamics.

Even though the stationary state is always featureless in our setting, the relaxation into that state is not trivial, due to the hydrodynamic long-time tails that emerge because of a redistribution of the conserved quantity during the dynamics~\cite{Wang_2025,Capizzi_2025,Matthies_2026}. This enables the system to relax into the same stationary state according to different power laws, depending on the specific initial state, see Fig.~\ref{fig:2},\ref{fig:3}. 

\textit{Results} -- In order to study this relaxation, we perform finite size numerical simulations of the dynamics of $\hat{\rho}_{A}$ for subsystems $A$ of a length of up to $l=6$ sites~\cite{Cirq_2025}. For the MFI model, we discretize time in steps of $dt=0.05$ and use a second order trotterization method~\cite{Hatano_2005}. We checked numerically that the results do not change qualitatively between $dt =0.1$ and $dt=0.05$ to rule out trotter errors. 

The QME has been discussed in terms of various different distance measures and witnesses in state space~\cite{Nielsen_Chuang_2010,Ares_2023}, and crossings thereof. We focus here on the trace distance $D_A$, the entanglement entropy $S_A$ and the entanglement asymmetry $\Delta S_{A,Q}$ with respect to the total magnetization. 

(i) The trace distance is defined as
\begin{equation}
    D_{A}(t) = \frac{1}{2} \Tr \vert \hat{\rho}_{A}(t) - \hat{\Lambda}_A \vert = \frac{1}{2} \sum_i \vert \lambda_{A,i}(t) - 1/2^l \vert,
\end{equation}
where $\lambda_{A,i}(t)$ are the eigenvalues of $\hat{\rho}_{A}(t)$. It fulfills all properties of a distance in state space and approaches $0$ if the system indeed thermalizes into the infinite temperature state. Our numerical simulations confirm that for both systems and for all considered initial states indeed upon increasing the system size and the simulated time, we obtain $D_A \to 0$, confirming that the system relaxes into the infinite temperature state. The finite size effects however strongly depend on the initial state, see Fig.~\ref{fig:3}. In both models, we observe the emergence of a power-law regime $D_{A} \sim t^{-1/z_\text{eff}}$ with an effective dynamical critical exponent $z_\text{eff}$ at intermediate times (see Fig.~\ref{fig:2}a),b)) $t_\text{short}<t<t_\text{long}$. This indicates an emergent hydrodynamic behavior~\cite{Spohn_1991}. $t_\text{short}=\mathcal{O}(N^0)$ is a non-universal short-time scale and $t_\text{long}=\mathcal{O}(N^z)$, where $z$ is the dynamical critical exponent of the model, is the time-scale at which finite size effects become important, see Fig.~\ref{fig:3}.

In the Floquet model (see Fig.~\ref{fig:2}b),~\ref{fig:3}a),c)), we observe that for $a=0$, $z_\text{eff} \approx \frac{2}{3}$, while for all other initial states, $z_\text{eff} \approx 2$ at late times. If $a\neq 0$ but small, we observe a transient regime where $z_\text{eff} \approx \frac{2}{3}$ before a crossover into $z_\text{eff} \approx 2$. This crossover gives rise to crossings of the trace distance between different initial states and, in particular, with the $a=0$ curve. We observe that the Mpemba crossing occurs later, the smaller $a$ is chosen, because the crossover in $z_\text{eff}$ occurs later. This indicates that there is a slowest decaying contribution to $D_A$, going to $0$ as $\sim t^{-1/2}$, which vanishes for $a \to 0$, and a sub-leading faster relaxing mode $\sim t^{-3/2}$. The fact that we can tune the leading algebraic decay to $0$ by $a \to 0$ gives rise to a robust and systematic appearance of Mpemba crossings in the hydrodynamic regime for the trace distance.

In the MFI model (see Fig.~\ref{fig:2}b),~\ref{fig:3}b),d)), there is an even larger set of different numerically found effective dynamical critical exponents. For $\theta=\pi/2$, we find $z_\text{eff}\approx 3/4$, for $\theta < \pi/2$ we obtain $z_\text{eff} \approx 1/4$ and for $\theta > \pi/2$ we have $z_\text{eff} \approx 3/2$. This demonstrates that even though for a typical initial state, the system behaves diffusively, $z=2$, specific initial states may show various different relaxation patterns~\cite{Banuls_2011}. This behavior can give rise to Mpemba crossings in the hydrodynamic regime, for instance comparing the relaxation from $\theta=\frac{3\pi}{8}$ and $\theta=\frac{5\pi}{8}$, see Fig.~\ref{fig:2}b).

(ii) The entanglement entropy is defined as
\begin{equation}
    S_{A}(t) = -\Tr \hat{\rho}_{A}(t) \log \hat{\rho}_{A}(t).
\end{equation}
Since $\hat{\rho}_{A} \to \hat{\Lambda}_A\sim \hat{1}$, the entanglement entropy approaches its maximal value at late times, $S_{A} \to l \log 2$ for all initial states. For that reason, we consider the missing entanglement to reach the maximum value, $\Delta S_{A}=l \log 2-S_{A} \to 0$ as a witness for the distance to the stationary state. We find in both models the same qualitative behavior as for the trace distance upon replacing $z_\text{eff} \to z_\text{eff}/2$, including the appearance of Mpemba crossings in the hydrodynamic regime, as shown in Fig.\ref{fig:2}b),c).

(iii) We also compute the entanglement asymmetry, as it has been recently highly relevant in the discussion of Mpemba effects~\cite{Murciano_2024,Yamashika_2024,Caceffo_2024,Fujimura_2025,Ares_2025,Yamashika_2025,Turkeshi_2024,Banerjee_2025,Chalas_2024,Calabrese_2026,Di_Giulio_2025}. It is defined as
\begin{equation}
    \Delta S_{A,Q}(t) = \Tr \hat{\rho}_{A}(t) (\log \hat{\rho}_{A}(t) - \log \hat{\rho}_{A,Q}(t)),
\end{equation}
where $\hat{\rho}_{A,Q}(t)=\sum_q \hat{\Pi}_q \hat{\rho}_{A}(t) \hat{\Pi}_q$ is the symmetrized reduced density matrix. $\hat{\Pi}_q$ are the projectors onto the eigenspaces with eigenvalue $q$ of an operator $\hat{Q}_{A}$ acting on subsystem $A$. In the literature, the case has been discussed where $\hat{Q}_A$ is a sum of local operators, such as $\hat{Q}_A=\sum_{j \in A} \hat{Z}_j$ for the magnetization. In particular, it was discussed in scenarios where $\hat{Q}=\sum_j \hat{Z}_j$ is a conserved quantity in the dynamics. Note that $[\hat{Q}_A, \hat{\Lambda}_A]$ for all symmetries $\hat{Q}$ since $\hat{\Lambda} \propto \hat{1}$. This implies that both dynamics restore all symmetries $\hat{Q}$ in a trivial way. Hence, we can study the relaxation of $\Delta S_{A,Q}$ for $\hat{Q}=\hat{Q}_\text{FL}$ for both models. In the MFI model the entanglement asymmetry behaves essentially like the entanglement entropy since magnetization is not conserved: We observe the same algebraic scaling behavior, and the same Mpemba crossings, see Fig.~\ref{fig:2}f). This is different for the Floquet model, where the chosen symmetry for which we compute the entanglement asymmetry is also a symmetry of the generator of the dynamics, not only of the stationary state, $[\hat{Q},\hat{U}_\text{FL}]=0$. Here, the decay of the entanglement entropy is much faster (but still algebraical), indicating that the slowest decaying modes do not appear in the entanglement asymmetry, as it is blind to the symmetry-compatible excitations. The modes that break the symmetry are decaying faster because they are sub-leading, as we will see in the hydrodynamic description below. We do not see Mpemba-crossings in our finite size numerical simulations here since the finite size correction to the stationary state becomes relevant very quickly due to the fast decay of the sub-leading terms, see Fig.~\ref{fig:2}e).

\textit{Hydrodynamic description} -- The numerical finding of emergent power-law relaxation upon increasing the system size indicates, that the relaxation is dominated by many-body effects rather than local relaxation. In order to understand the QME in our setting, we therefore have to resort to an effective many-body theory. Both models have a conserved quantity, whose correlations need to be transported through the whole system, which gives rise to hydrodynamic (power-law) long time tails in correlation functions~\cite{Spohn_1991}. Therefore, hydrodynamics is a promising candidate for such an effective many-body theory. For the Floquet model, this approach allows us to provide a full explanation of the numerical observations, following Ref.~\cite{Matthies_2026}. Since there is only one conserved quantity, magnetization, we can describe $\hat{Z}_i$ by a single diffusive mode $m(x,t)$, and all other degrees of freedom by a stochastic noise~\cite{Spohn_1991}. Their dynamics is governed by a Langevin equation, entirely fixed by the symmetries of the coarse-grained system. This allows to compute all moments of $m(x,t)$ averaged over noise realizations, which can be associated with expectation values of local Pauli-strings $\hat{P}$, as demonstrated in Ref.~\cite{Matthies_2026}. This is useful for our purposes, as we can represent the reduced density operator as
\begin{equation}
    \hat{\rho}_A = \frac{1}{2^l} \sum_{\hat{P} \in \mathcal{P}(A)} \langle \hat{P} \rangle \hat{P},
\end{equation}
where $\mathcal{P}(A)$ is the set of all Pauli-strings on $A$, i.e. tensor products of $\hat{1},\hat{X},\hat{Y},\hat{Z}$ on $A$. Since the system relaxes to the infinite temperature state, for all $\hat{P}\neq \hat{1}$, the expectation value vanishes at late times, $\langle \hat{P} \rangle_t\to 0$. Some of them are protected by symmetry. For those to relax, the conserved quantity needs to be transported through the system diffusively, leading to algebraic decay. This is most clear for operators only containing $\hat{Z}$ such as $\langle \hat{Z}_{i} \hat{Z}_{i+1} \rangle_t \sim \overline{m^2(x,t)}$. Besides that, there are operators involving $\hat{X}$ and $\hat{Y}$, which also commute with $\hat{Q}$ and are therefore protected as well, for instance $[\hat{X}_i \hat{X}_{i+1}+\hat{Y}_i \hat{Y}_{i+1},\hat{Q}] = 0$. These operators have to be associated with derivatives of the magnetization, $\hat{X}_i \hat{X}_{i+1} \sim (\nabla m)^2$ which implies faster but still algebraic relaxation, making them sub-leading at late times. Finally, there are operators not protected by the symmetry, which are expected to decay exponentially.

If we now initialize the system in a generic state, all initial Pauli-string expectation values will be non-vanishing, and contribute to the relaxation. On a coarse grained level, such an initial state has $\overline{m(x)}=0$ such that the leading contribution comes from $\overline{m(x,t)m(y,t)}\sim t^{-1/2}$~\cite{Matthies_2026,Spohn_1991} i.e. $\hat{P}=\hat{Z}_i \hat{Z}_j$ for $i\neq j \in A$. These correlations therefore decay as $\sim t^{-1/2}$ which allows us to compute the leading contribution to the density operator
\begin{equation}
    \hat{\rho}_{A}(t) \simeq \hat{\Lambda}_A + \frac{1}{2^l}\sum_{i \neq j \in A} \langle \hat{Z}_i \hat{Z}_j \rangle_t \hat{Z}_{i} \hat{Z}_j . 
\end{equation}
For the trace distance, this implies 
\begin{equation}
    D_{A}(t) \simeq \frac{1}{2^l} \Tr  \Bigg\vert \sum_{i\neq j \in A} \langle \hat{Z}_i \hat{Z}_j \rangle_t \hat{Z}_{i} \hat{Z}_j \Bigg\vert  \sim t^{-1/2}.
\end{equation}
Similarly, for the entanglement entropy, we obtain
\begin{equation}
    \Delta S_{A}(t) \simeq \frac{1}{2^{l-1}} \sum_{i\neq j \in A} \langle \hat{Z}_i \hat{Z}_j \rangle_t^2 \sim t^{-1},
\end{equation}
where we used that in the leading order expansion of the logarithm, $\Tr \hat{Z}_i \hat{Z}_j=0$ for $i\neq j$.
The entanglement asymmetry on the other hand only depends on Pauli-strings that do not commute with all $\hat{\Pi}_q$. Therefore, all operators that only contain $\hat{Z}$ drop out and the leading contribution comes from operators such as $\hat{X}_i \hat{X}_{i+1}\sim (\nabla m)^2\sim t^{-3/2}$. Just as for the entanglement, the leading contribution in an expansion around the infinite temperature state only appears at second order such that
\begin{equation}
    \Delta S_{A,Q}(t) \sim (\nabla m)^4 \sim t^{-3}.
\end{equation}
The predictions for $D_A$ and $\Delta S_A$ are compatible with the results for the late time dynamics of the Floquet model for $a \neq 0$, as shown in Fig.~\ref{fig:2}a),c). For the asymmetry, strong finite size effects hide the correct power law at late times, see Fig.~\ref{fig:2}f).

Why is the relaxation faster for $a=0$? In fact, we have considered very specific initial states, and not sampled them from the Haar measure. In particular, for $a=0$, the $+$ polarized initial state, we have $\langle \hat{Q}^2 \rangle_0=\Tr \hat{Q}^2 \hat{\Lambda}$. Since $[\hat{Q},\hat{U}_\text{FL}]=0$, this remains true for all times. This implies that the coarse-grained fluctuations in the initial and the final state are exactly the same, and they do not have to be transported diffusively. For that reason, $\overline{m(x,t) m(y,t)}\sim t^{-3/2}$ is dominated by the first sub-leading contribution, as discussed in Ref.~\cite{Matthies_2026}. We conclude that $\langle \hat{Z}_i \hat{Z}_j \rangle_t \sim t^{-3/2}$ such that $D_{A}\sim t^{-3/2}$ and $\Delta S_{A} \sim t^{-3}$ matching our numerical findings shown in Fig.~\ref{fig:2}a),c),~\ref{fig:3}c). For the entanglement asymmetry, we also obtain an additional $t^{-1}$ factor for the decay of $\hat{X}_i \hat{X}_{i+1}+\hat{Y}_i \hat{Y}_{i+1}\sim t^{-5/2}$ in the $a=0$ initial state, leading to $\Delta S_{A,Q}\sim t^{-5}$ compatible with our findings in Fig.~\ref{fig:2}c). However, we do not see slower decay for $a \neq 0$ due to finite size effects in this observable. To summarize, the hydrodynamic description predicts the leading late-time behavior
\begin{align}
    D_{A} &\sim \begin{cases}
        t^{-1/2} & a \neq 0 \\
        t^{-3/2} & a = 0
    \end{cases} ,\\
    \Delta S_{A} &\sim \begin{cases}
        t^{-1} & a \neq 0 \\
        t^{-3} & a = 0
    \end{cases} ,\\
    \Delta S_{A,Q} &\sim \begin{cases}
        t^{-3} & a \neq 0 \\
        t^{-5} & a = 0
    \end{cases} .
\end{align}

\textit{Conclusion} -- Chaotic quantum systems with conservation laws can thermalize unconventionally fast for some special initial states. Other states that relax slowly can be either closer or further from the stationary state. If we found such a state at larger distance, we can find another conventionally relaxing state closer to the stationary state by evolving the state until it is closer than the fast relaxing state. Taking this state as a new initial state, we can then observe the Mpemba crossing. Hence, the crucial requirement for the observation of the QME is indeed the appearance of unconventional relaxation behavior for specific initial states that differ from generic initial states. We demonstrated this general principle for states thermalizing into a trivial infinite temperature state. This raises the question if the mechanism is transferable to relaxation into structured states at finite or even low temperature, with potential application to enable fast state preparation. While the Floquet model can be understood in the hydrodynamic picture, the relaxation of highly symmetric initial states in the MFI model is far less clear, as was also noted in Ref.~\cite{Banuls_2011}. We observe both sub-diffusive as well as super-diffusive relaxation with unconventional exponents, depending on the initial state. These findings are consistent with earlier indications that this model can exhibit anomalous thermalization dynamics~\cite{Banuls_2011, Kormos_2016, Mazza_2019, Jiang_2026}. A comprehensive understanding of its emergent hydrodynamic behaviour remains an open problem and is left for future investigation.

Besides the case of faster relaxation of certain initial states, as discussed here, there are phenomena such as approximate many-body scars~\cite{Turner_2018}, providing unconventionally slow thermalization of some initial states, while others thermalize quickly. It will be interesting to explore models hosting these effects to find further potential origins for QMEs in chaotic many-body quantum systems. 

\textit{Acknowledgements} -- We thank Gerald Fux, Emanuele Tirrito and Xhek Turkeshi for useful discussions. This work was supported by the
European Union (ERC, RAVE, Grant No. 101053159).  SP acknowledges funding by the Deutsche Forschungsgemeinschaft (DFG, German Research Foundation) under Projektnummer 277101999 - TRR 183 (project B02), and under Germany's Excellence Strategy - Cluster of Excellence Matter and Light for Quantum Computing (ML4Q) EXC 2004/1 - 390534769. Raw data and code to generate the figures is available on Zenodo~\cite{Zenodo}.

\bibliography{Bib}

\end{document}